# Soliton Distillation in Fiber Lasers

Yutian Wang,[1] Songnian Fu,[1] Chi Zhang,[1] Xiahui Tang,[1] Jian Kong,[2] and Luming Zhao[1,2,*]

[1]*Wuhan National Laboratory for Optoelectronics, and School of Optical and Electronic Information, Huazhong University of Science and Technology, Wuhan, 430074, China*
[2]*Kunshan Shunke Laser Technology Co., Ltd, Suzhou, 215347, China*
*Corresponding author: lmzhao@hust.edu.cn*

**Pure solitons are for the first time distilled from the resonant continuous wave (CW) background in a fiber laser by utilizing nonlinear Fourier transform (NFT). It is identified that the soliton and the resonant CW background have different eigenvalue distributions in the nonlinear frequency domain. Similar to water distillation, we propose the approach of soliton distillation, by making NFT on a steady pulse generated from a fiber laser, then filtering out the eigenvalues of the resonant CW background in the nonlinear frequency domain, and finally recovering the soliton by inverse NFT (INFT). Simulation results verify that the soliton can be distinguished from the resonant CW background in the nonlinear frequency domain and pure solitons can be obtained by INFT.**

The term 'soliton' refers to the localized solution of integral nonlinear systems [1]. Solitons are the balanced product of diffraction and nonlinearity, where they can propagate without distortion in the medium. In optics, solitons can be generated by a cancellation of nonlinear and dispersive effects in the medium. Due to the anomalous dispersion of fiber, the existence of solitons in optical fibers was theoretically predicted [2] and experimentally verified [3], which greatly promoted the exploration of solitons in nonlinear systems. Due to the natural information representor of "1" for binary communication, ultrashort pulse duration and distortion-less transmission, solitons are considered as a perfect candidate for high-speed telecommunications in fiber optical transmissions [4]. Strictly speaking, solitons only exist in conservative systems, for example, a system that can be governed by the nonlinear Schrödinger equation (NLSE). With a loose definition, ultrashort pulses generated in a fiber laser can be considered as solitons, since the pulse repeats itself at a fixed position in the fiber laser under the condition of steady state. Solitons generated in fiber lasers can be considered as an averaged soliton when all cavity parameters are distributed [5]. To achieve soliton generation in fiber lasers, various passive mode-locking techniques are applied [6-8]. Different from pulse propagation in a fiber where the system is conservative under the assumption of propagation without loss, a fiber laser is a dissipative system where gain and loss affect the pulse generation while simultaneously it is a periodic boundary system.

Independent of detailed mode-locking mechanism, solitons generated in fiber lasers experience periodical disturbances, such as gain and loss, and release the excess energy as dispersive waves. If the phase difference between the soliton and the dispersive wave is multiples of $2\pi$, they will interfere with each other and Kelly sidebands will symmetrically appear on the spectrum [9]. In addition, incoherent sidebands may appear on the spectrum too [10]. Here, both Kelly sidebands and incoherent sidebands are collectively referred to the resonant continuous wave (CW) background. As Kelly sidebands coherently interfere with solitons, it is challenging to get rid of them. People always used an external spectral filter to filter out the CW component before amplifying the soliton for specific applications [11]. However, spectral filtering simultaneously changes solitons. So far, no method has been proposed to separate solitons from the parasitic resonant CW background. Alternatively, people prefer to use the appearance of Kelly sidebands as a criterion to verify the generation of soliton [12,13]. The true optical spectrum of a pure soliton generated from a fiber laser is so far unknown. It is interesting to know the details of a pure soliton, instead of a composite of a soliton and the resonant CW background. The coherent and incoherent spectral sidebands of the composite spectrum have been identified by using the dispersive Fourier transform (DFT) technique [10]. To separate a soliton from the mixture of solitons and the resonant CW background from a fiber laser can answer a fundamental distress that what is the real difference between a soliton from a conservative system and a dissipative system. The implementation of pure solitons would also rule out background induced soliton interaction [14], which will greatly simplify analysis of soliton interaction in nonlinear systems.

Recently, an elegant method of nonlinear Fourier transform (NFT) has attracted worldwide research interests [15], which can transform the signal into the nonlinear spectrum, including continuous spectrum and discrete spectrum, where eigenvalue lies on the upper-half complex plane. In such an approach, information is encoded into the nonlinear spectrum of signal, which can effectively address the nonlinear transmission impairments arising in standard single mode fiber (SSMF) [16]. Meanwhile, as a method to obtain analytic solutions to the NLSE, NFT can also be used to analyze signals in optical fibers, for instance, the rogue waves [17]. Theoretically, for a pure soliton, its nonlinear spectrum only contains discrete spectrum and the eigenvalues in the discrete spectrum correspond to characteristics of the soliton, whose real and imaginary parts correspond to the frequency and amplitude of the soliton, respectively. Such NFT methodology provides a new viewpoint on the physics of laser dynamics. Recently, the use of the NFT has been proposed for the investigation of laser radiation [18-20], indicating the capability to characterize the ultrashort pulse in the nonlinear frequency domain. Therefore, it is possible to separate solitons and the resonant CW background according to different eigenvalue distributions.

In this paper, we investigate the separation between solitons and the resonant CW background in a mode-locked fiber laser, and analyze the dynamic of pulses in the nonlinear spectrum after NFT. Since, in the nonlinear frequency domain, the soliton and the resonant CW background have different eigenvalue distributions, we can separate the soliton from the resonant CW background. After eigenvalue separation in the nonlinear frequency domain, we are able to recover the temporal waveform of a pure soliton by inverse nonlinear Fourier transform (INFT). The whole procedure is similar to water distillation. Therefore, we call it as soliton distillation.

The nonlinear spectrum of a given pulse $q(t)$ is defined [21]

$$\tilde{q}_c(\lambda)=b(\lambda)/a(\lambda), \lambda \in R$$
$$\tilde{q}_d(\lambda_n)=b(\lambda_n)/a'(\lambda_n), \lambda_n \in C^+ \quad (1)$$

where $a(\lambda)$ and $b(\lambda)$ are scattering data, $\tilde{q}_c(\lambda)$ is the continuous part of nonlinear spectrum and $\tilde{q}_d(\lambda_n)$ is its discrete part, while $a'(\lambda)=da(\lambda)/d\lambda$ and $\lambda_n$ is the root of $a(\lambda)$.

Thus, by analyzing the nonlinear spectrum of the evolving pulse, we can readily segregate the continuous spectrum (representing the dispersive component) from the complex eigenvalue spectrum (referring to the invariant or coherent solitary components). When calculating the nonlinear spectrum, different sets of normalization parameters will lead to different energy distributions between nonlinear spectral components. Since the nonlinear spectrum of a soliton only contains single discrete eigenvalue, its temporal waveform is [22]

$$q(t) = -2j\lambda_I e^{-j\angle\tilde{q}(\lambda_n)} \text{sech}(2\lambda_I(t-t_0))e^{-2j\lambda_R t}. \quad (2)$$

where $\lambda_R$ and $\lambda_I$ are the real and imaginary parts of eigenvalue $\lambda_n$, $\angle\tilde{q}(\lambda_n)$ is the spectrum phase, and $t_0$ is the time center associated with $\lambda_I$ and spectrum amplitude. As shown in Eq. (2), the eigenvalue $\lambda_n$ specifies the soliton parameters with an amplitude of $2\lambda_I$ and a frequency of $2\lambda_R$. Thus, the soliton can be characterized by its discrete eigenvalue.

We apply the NFT methodology to obtain the eigenvalues of pulses from a passively mode-locked fiber laser, after traditional coherent detection. A standard fiber laser mode locked by the nonlinear polarization rotation is built up to generate pulses [23], as shown in Fig. 1. All anomalous dispersion cavity design is achieved by using a segment of 3m erbium-doped fiber (EDF) with a dispersion parameter of 23 ps/km/nm, in order to provide sufficient gain for laser emission. The other fibers are SSMFs and the total cavity length is 6 m. A polarization controller (PC) is employed to adjust the polarization of the propagating light and a polarization-dependent isolator guarantees the unidirectional operation. The laser outputs are hyperbolic secant pulses with a width of ~ 180 fs, center wavelength of 1560 nm, and a spectral bandwidth of ~ 8.8 nm. The output ratio is 10%.

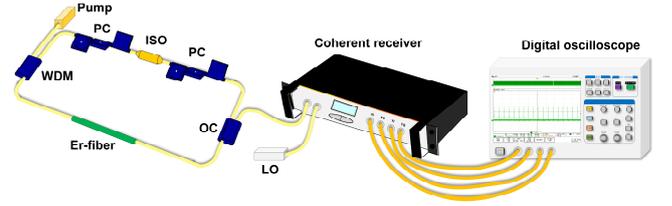

Fig. 1. Fiber laser and proposed measurement setup. ISO — isolator; PC — polarization controller; WDM — wavelength division multiplexing; OC — output coupler; LO — local oscillator.

For the ease of implementing traditional coherent detection, the output is mixed with a semiconductor laser source having a narrow linewidth of 100 kHz act as local oscillator (LO), with the help of 90º hybrid, which gives rise to the generation of in-phase (I) and the quadrature (Q) components. The optical-to-electrical conversion is realized by a pair of balanced photodetectors (BPDs) and four electrical outputs are digitalized by a high-speed oscilloscope. Finally, the NFT determined eigenvalue can be calculated from the full-field information of the pulse.

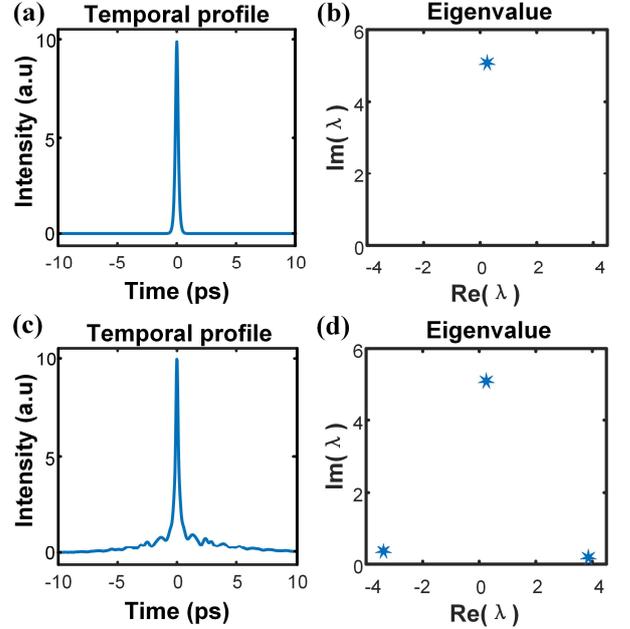

Fig. 2. (a) Temporal profile of a theoretical soliton, (b) eigenvalue of a theoretical soliton, (c) the temporal pulse profile arising in a mode-locked laser and (d) its eigenvalues distribution.

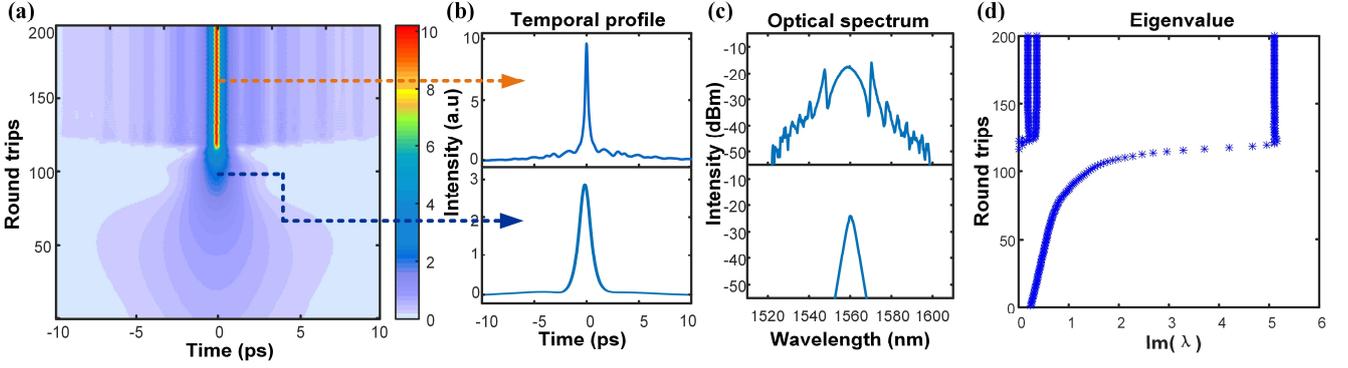

Fig. 3. NFT data evolution obtained from the measurement of the fiber laser field: (a) real-time spatial-temporal dynamics of laser evolution from full-field measurements; (b), (c) temporal waveform and spectrum of laser pulse before and after the generation of soliton, respectively; (d) the evolution of the imaginary parts of eigenvalues.

During the coherent detection, both the frequency offset (FO) between the pulse and LO and the linewidth of LO will introduce the penalty to the characterization results. We set the FO as 1 GHz in the simulation to mitigated both FO and phase noise effects the same as that for the nonlinear frequency division multiplexing (NFDM) transmission [24,25]. Meanwhile, in order to improve the NFT accuracy, we resample the data acquired by the oscilloscope so that each pulse has at least 1024 sampling points.

Fig. 2(a) shows a theoretical soliton described by Eq. (2) and with a pulse duration of 180 fs. Fig. 2(b) shows its NFT eigenvalue. A typical output of the passively mode-locked fiber laser and its corresponding NFT eigenvalues are shown in Figs. 2(c) and 2(d). As shown in Fig. 2(d), the eigenvalue corresponding to the soliton has large imaginary parts and almost zero real parts, which indicates that the soliton has a zero velocity. The imaginary part of the soliton eigenvalue is related to the laser pulse amplitude. Meanwhile, some eigenvalues have non-zero real parts and relatively small imaginary parts, indicating of non-zero velocities of the corresponding temporal features of the resonant CW background. The real parts of CW eigenvalues are related to its frequency. It is straightforward that the eigenvalues of the pulse from the fiber laser is different from that of the theoretical soliton, which can be attributed to the coexisting resonant CW background. Since the small $\lambda_I$ corresponds to the resonant CW background, it is possible to separate pure soliton and the resonant CW background based on NFT.

Then, we analyze the buildup dynamics of solitons in the name of nonlinear Fourier spectrum. To calculate the nonlinear spectrum from these full-field dynamics, we normalize time to a scale $T_s = 1 ps$, which approximately contains 99% of the signal energy averaged over many round trips. The amplitude is also normalized with a scale $Q_s$, which is the squared root of the power of a hyperbolic secant signal with a time-width equal to a time window containing 99% of the signal energy. The dynamic process of laser pulse is shown in Fig. 3. The intensity variation of optical field is accompanied by the variation of eigenvalue spatial distribution. The intensity evolution of laser pulse is shown in Fig. 3(a), where different colors correspond to different intensities. Figs. 3(b) and 3(c) show the temporal waveform and optical spectrum before and after the rising of the soliton, respectively. When the soliton is generated, the resonant CW background appears, and Kelly sidebands symmetrically exist in the optical spectrum. Fig. 3(d) shows the imaginary part evolution of eigenvalues versus roundtrips. Accompanying the soliton generation, there exist some eigenvalues with smaller imaginary parts in the nonlinear spectrum, corresponding to the resonant CW background.

In nonlinear frequency domain, soliton and the resonant CW background have different eigenvalue distributions, which can be used to distinguish different components of the pulse. Naturally, if the eigenvalues of the resonant CW background are removed in the nonlinear frequency domain and only the soliton component remains, we can recover the pure soliton in the time domain through the INFT. Figs. 4(a) and 4(b) show the temporal waveform and optical spectrum, respectively, where the differences can be clearly observed among the pulse from the fiber laser, pulse after soliton distillation, and theoretical soliton. The theoretical soliton is constructed according to the amplitude and pulse-width of the pulse from the fiber laser $q(t) = A\ sech(1.763 \times (t - t_0)/\tau)$, where $A$, $t_0$ and $\tau$ are the amplitude, central time and pulse-width of the pulse from the fiber laser, respectively. The temporal waveform and optical spectrum of the pulse after the soliton distillation are almost the same as the theoretical soliton. Thus, our proposed soliton distillation based on the NFT is successfully verified.

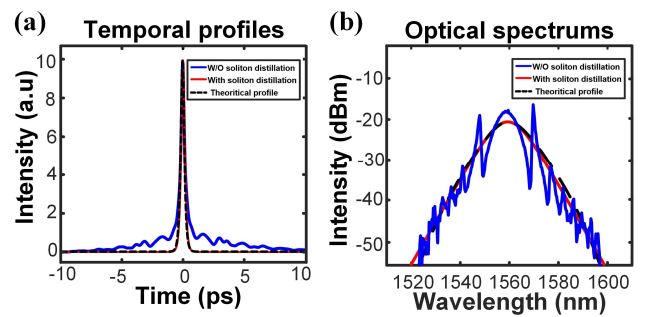

Fig. 4. (a) Temporal waveforms of laser pulse, filtered soliton and structure soliton and (b) optical spectral of them.

The coexistence of resonant CW background will significantly affect pulse interaction when multiple pulses appears. With the help of soliton distillation, we can investigate the case of soliton interaction without the resonant CW background. A typical state of double pulses could be easily obtained from the fiber laser with appropriate parameter setting. The temporal waveform, optical spectrum and corresponding eigenvalues of a double-pulse state are shown in Figs. 5(a), 5(b) and 5(c), respectively. Due to the resonant CW background, there are obvious Kelly sidebands in the optical spectrum and some eigenvalues with large imaginary parts and almost zero real parts in the nonlinear spectrum. Similarly, we select the eigenvalues corresponding to solitons and reconstruct the temporal waveform by the INFT. The reconstructed soliton pair after soliton distillation is shown in Fig. 5(d). It is found that the soliton separation always reduced after soliton distillation. It verified that the resonant CW background actually buffers solitons and affects the performance of coexisting soliton pair.

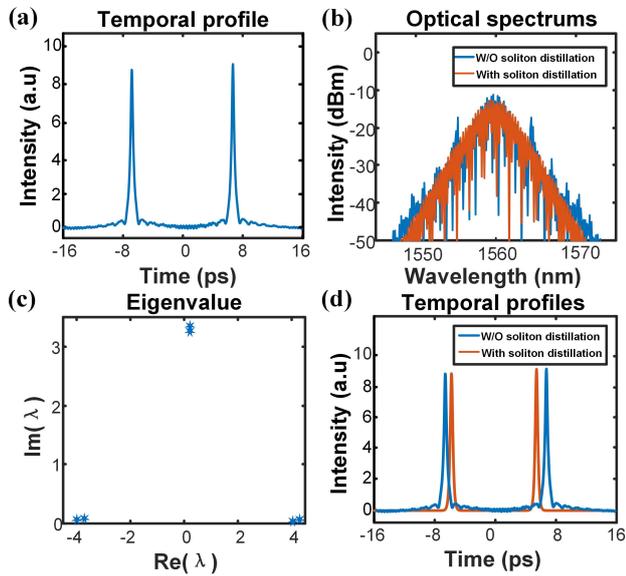

Fig. 5. (a) Temporal waveform of double pulses from the fiber laser, (b) spectra of laser pulse, filtered soliton and structure soliton, (c) eigenvalues of laser pulse, and (d) waveforms of laser pulse, filtered soliton and structure soliton.

In conclusion, we first distill pure solitons from the resonant CW background in a fiber laser by utilizing NFT. In the nonlinear frequency domain, the soliton and the resonant CW background have different eigenvalue distributions. After filtering out the eigenvalues of the resonant CW background in the nonlinear frequency domain, we can recover pure soliton by INFT. Simulation results verify that the solitons can be distinguished from the resonant CW background in the nonlinear frequency domain and pure solitons can be obtained by INFT, which is of great help for the fundamental understanding of solitons. Soliton distillation also provide a method to study soliton interaction without the resonant CW background. NFT pave a new way for soliton studying in dissipative nonlinear systems.

**Funding.** National key R&D Program of China (2018YFB1801002); Fundamental Research Funds for the Central Universities (HUST: 2020kfyXJJS007); National Natural Science Foundation of China (61875061, 11674133).

**Acknowledgement.** The concept of soliton distillation was partially inspired by my 7-years-old daughter, Yuzhen Ithaca Zhao, during the Covid-19 lockdown.